\documentclass[%
 reprint,
 amsmath,amssymb,
 aps,
]{revtex4-1}
\usepackage{changepage}
\usepackage{braket}
\usepackage{graphicx}
\usepackage{dcolumn}
\usepackage{bm}
\usepackage[caption=false]{subfig}
\usepackage{floatrow}
\usepackage{natbib}

\begin{document}

\preprint{APS/123-QED}

\title{Nonlinear Time Domain Spectroscopy Near a Band Inversion}

\author{Zachariah Addison}%
\affiliation{Department of Physics and Astronomy,
University of Pennsylvania,
Philadelphia, PA 19104, USA}
\author{E. J. Mele}
\affiliation{Department of Physics and Astronomy,
University of Pennsylvania,
Philadelphia, PA 19104, USA}

\date{\today}

\begin{abstract}
We develop a theory for the nonlinear time domain response of a Weyl semimetal driven by an ultrafast optical pulse. At quadratic order in the driving field we find that the response near a band inversion transition contains a coherent oscillating component proportional field intensity and with a frequency that can be tuned over a wide range (from THz to the near IR) selected by the chemical potential. We illustrate the effect by calculating the induced current as a function of time for a model of a parity broken Weyl semimetal where large Berry curvature near the band inversion transition promotes this nonlinear response.
\end{abstract}
\maketitle

\section{Introduction}

Electrodynamic constitutive relations in the bulk of a crystal can be used to interrogate the quantum geometric character of its band structure. When formulated in the frequency domain, nonlinear responses quadratic in the driving fields have been identified as probes of the distribuion of momentum space Berry curvature  in systems with broken inversion symmetry \cite{keller1985theoretical,pedersen2015intraband,morimoto2016topological} or broken mirror symmetries \cite{de2017quantized,ji2019spatially}.  Generally, point nodes in a band structure are associated with regions of ${\mathbf{k}}$ space with enhanced Berry curvatures and large interband matrix elements both of which are known to promote nonlinearities in the frequency dependent optical response.

In Weyl semimetals and other band inverted systems large Berry curvatures are often found in the momentum space region close to a band contact point and below a small Lifshitz energy scale. Above this energy, iso-energy surfaces in the band structure do not close around  isolated singularities which prevents a direct identification of the band topology. Frequency domain spectroscopy below this Lifshitz scale typically requires interrogation at infrared or THz frequencies.  In this work we consider instead the manifestations in  nonlinear time domain spectroscopy. We find that excitation with a  ultrafast high intensity pulse produces a nonlinear response; one part of which describes a coherent oscillation of the induced currents at a tunable frequency set by the chemical potential. We demonstrate that a nonlinear response near a Pauli blocked threshold selects a frequency-tunable response from a broad band source.  This generically occurs for a narrow gap system near a band inversion transition, and the strength of the nonlinearity can be enhanced by the large matrix elements that can be associated with a topological transition in the band structure.

This application is a variant of a class of well studied phenomena at higher frequencies where one drives electronic motion with ultrafast high-intensity electromagnetic pulses.  This family of novel nonlinear responses has been studied in interference measurements with atto-second electric fields \cite{dudovich2006measuring,schultze2013controlling,krausz2014attosecond, lucchini2016attosecond} and has been observed in  the petahertz dynamics in semiconductors like gallium nitride using few cycle near-infrared electromagnetic pulses \cite{mashiko2016petahertz}.  Similarly in silicon the transfer of electrons from valence band states to conduction band states have been observed in the extreme ultraviolet absorption spectrum using attosecond interferometry \cite{schultze2014attosecond}.  Sub-cycle motion of electrons in driven terahertz phase locked pulses have been studied in real time by studying the interband quantum interference of electrons in Bloch states far below the Fermi energy \cite{golde2008high, hohenleutner2015real}.

Here we develop the theory for time domain dynamics for electrons in semimetals and narrow gap semiconductors near a band inversion transition.  We study nonlinearities to quadratic order in the driving fields and show that dipole mediated transitions between states on the fermi surface and above the fermi surface can generate a coherent current oscillation with a frequency tuned by the chemical potential.  We find that all other allowed electronic transitions add incoherently and lead to non-oscillatory current generation.  The frequency of current generated by the ultrafast pulses is set by the energy scale between the Fermi energy and energy of band states above the Fermi surface.

Charge currents in materials produced by nonlinear coupling to optical fields is often studied in the frequency domain \cite{sipe2000second, parker2019diagrammatic, PhysRevB.97.235446}.  These theoretical treatments are useful when investigating the low frequency charge currents that are produced by the nonlinear downconversion of optical fields.  For example at second order in a perturbing electric field, DC currents like the shift and injection current are generated by electric fields with a single frequency \cite{kral2000quantum, nastos2006optical, bieler2006simultaneous, young2012first, tan2016shift,  cook2017design}.  Conversely, processes like second harmonic generation produce currents double the frequency of the driving electric field \cite{keller1985theoretical,pedersen2015intraband, morimoto2016topological}.

Here we are interested in currents generated by  ultrafast electric field pulses.  The simple processes that lead to shift, injection, and second harmonic currents are difficult to isolate in this limit.  Instead we directly study the currents in a time domain formulation and isolate second order processes by their dependence on the field intensity.  We compute the time dependent quantum density matrix to second order in a perturbing electric field and trace with the current operator to obtain the time dependent induced current density.   We illustrate the phenomena using a simple model for a time reversal symmetric, but inversion breaking 2D semiconductor to calculate these induced currents.  We find oscillating currents at frequencies selected by a Pauli threshold set by the electronic doping level of the semiconductor.

\section{Quantum Kinetic Equation for Bloch Electrons in an External Electric Field}

In order to calculate the time dependent charge current to quadratic order in a perturbing electromagnetic field we first solve for the electronic charge density to second order in an external electric field by iteratively solving the quantum kinetic equation for the density matrix $\hat{\rho}$.  This equation derives from the von Neumann equation that describes the time evolution of this quantum operator \cite{von1927wahrscheinlichkeitstheoretischer}.

\begin{equation}
\dfrac{d\hat{\rho}(t)}{dt}=-\dfrac{i}{\hbar}[\hat{H}(t),\hat{\rho}(t)]
\end{equation}

\noindent
Before application of a perturbing electromagnetic field the unperturbed hamiltonian $\hat{H}_0$ has eigenstates that are crystalline Bloch modes whose energy $\varepsilon_n(\bm{k})$ is index by the states crystal momentum $\bm{k}$ and band $n$ and whose periodic part we denote by the ket $\ket{u_n(\bm{k})}$.  We can write the von Neumann equation in this unperturb basis and denote matrix elements of the density matrix as $\rho_{nm}(\bm{k},t)=\bra{u_n(\bm{k})}\hat{\rho}(t)\ket{u_m(\bm{k})}$.  Here we will consider spatially homogeneous perturbing fields $\bm{E}(\bm{r},t)\rightarrow \bm{E}(t)$ that only couples Bloch electrons with the same Bloch wavevector $\bm{k}$ such that the perturbed density matrix is diagonal in crystal momentum $\bm{k}$: $\bra{u_n(\bm{k})}\hat{\rho}\ket{u_m(\bm{k}')}=\delta_{\bm{k},\bm{k}'}\bra{u_n(\bm{k})}\hat{\rho}\ket{u_m(\bm{k})}$.  Here we treat the coupling of the electromagnetic field to fermionic matter in the electronic dipole approximation $\hat{H}_{int}=e\bm{E}(t)\cdot\hat{\bm{r}}$ \cite{PhysRevB.96.035431}.  The von Neumann equation can than be written as

\begin{equation}\label{masteq}
\dfrac{d\hat{\rho}(t)}{dt}=-\dfrac{i}{\hbar}[\hat{H}_0(t),\hat{\rho}(t)]-\dfrac{i}{\hbar}[e\hat{\bm{r}}\cdot\bm{E}(t),\hat{\rho}(t)]
\end{equation}

\noindent
In the Bloch basis $\bra{u_n(\bm{k})}\hat{H}_0\ket{u_m(\bm{k})}=\delta_{nm}\varepsilon_n(\bm{k})$ and $\hat{\bm{r}}$ takes the representation $i\partial_{\bm{k}}$ \cite{blount1962advances}.  Substitution into equation \ref{masteq} leads to the quantum kinetic equation for the density matrix written in the Bloch basis and perturbed by a time dependent homogenous external electric field \cite{sekine2017quantum}.

\begin{widetext}
\small
\begin{equation}\label{mastereq3}
\dfrac{\partial \rho_{nm}(\bm{k},t)}{\partial t}+\bigg(\dfrac{i}{\hbar}(\varepsilon_n(\bm{k})-\varepsilon_m(\bm{k}))+\dfrac{1}{\tau}\bigg)\rho_{nm}(\bm{k},t)-\dfrac{\delta_{nm}f^T_n(\bm{k},\mu)}{\tau}=\sum_{i,l}\dfrac{eE_i(t)}{\hbar}\bigg(\partial_{k_i}\rho_{nm}(\bm{k},t)-i(R^i_{nl}(\bm{k})\rho_{lm}(\bm{k},t)-\rho_{nl}(\bm{k},t)R^i_{lm}(\bm{k}))\bigg)
\end{equation}
\normalsize
\end{widetext}

\noindent
Here $f^T_n(\bm{k},\mu)$ is the fermi occupation function that depends both on the temperature $T$ and chemical potential  $\mu$ of the system, $R^i_{nm}(\bm{k})=\bra{u_n(\bm{k})}i\partial_{k_i}\ket{u_m(\bm{k})}$ are the matrix elements of the dipole operator, and $\tau$ is a phenomenological relaxation constant arising from other electronic interactions.  As will be shown this constant will set the timescale for the system to return to its unperturbed equilibrium configuration.

\subsection{Gauge Covariance}

The dynamics of the system, like the charge density and current density, should be invariant under gauge transformations of the Bloch functions of the form $\ket{u_n(\bm{k})}\rightarrow e^{i\theta_n(\bm{k})} \ket{u_n(\bm{k})}$ for all $n$.  As such the quantum kinetic equation for the quantum density matrix should remain covariant under such a transformation.  The matrix elements of both the density operator $\rho_{nm}(\bm{k},t)$ and dipole operertor $\bm{R}_{nm}(\bm{k})$ are changed by the gauge transformations $\ket{u_n(\bm{k})}\rightarrow e^{i\theta_n(\bm{k})} \ket{u_n(\bm{k})}$ via

\begin{align}
\rho_{nm}(\bm{k},t)&\rightarrow e^{i(\theta_m(\bm{k})-\theta_n(\bm{k}))}\rho_{nm}(\bm{k},t)  \\ \nonumber \\
\bm{R}_{nm}(\bm{k})&\rightarrow e^{i(\theta_m(\bm{k})-\theta_n(\bm{k}))}\bm{R}_{nm}(\bm{k})+\delta_{nm}i\bm{\nabla}\theta_m(\bm{k})
\end{align}

\noindent
The left hand side of equation \ref{mastereq3}  under this gauge transformation is simply multiplied by the phase $e^{i(\theta_m(\bm{k})-\theta_n(\bm{k}))}$, while elements in the right hand side of equation \ref{mastereq3} transform as

\begin{align}
\partial_{k_i}\rho_{nm}(\bm{k},t)&\rightarrow e^{i(\theta_m(\bm{k})-\theta_n(\bm{k}))} \bigg(\partial_{k_i}\rho_{nm}(\bm{k},t)  \nonumber\\
&+i\rho_{nm}(\bm{k},t)(\partial_{k_i}\theta_m(\bm{k})-\partial_{k_i}\theta_n(\bm{k})) \bigg)  \\
\sum_{l}R^i_{nl}(\bm{k})\rho_{lm}(\bm{k},t)&\rightarrow e^{i(\theta_m(\bm{k})-\theta_n(\bm{k}))} \bigg(\sum_lR^i_{nl}(\bm{k})\rho_{lm}(\bm{k},t) \nonumber \\
&-i\partial_i \theta_n(\bm{k})\rho_{nm}(\bm{k},t)\bigg)
\end{align}

\noindent
Combining the above results demonstrates that the right hand side of equation \ref{mastereq3} is also simply multiplied by the phase $e^{i(\theta_m(\bm{k})-\theta_n(\bm{k}))}$ under this type of gauge transformation, implying that the quantum kinetic equation for the density matrix is gauge covariant.  The solutions to the quantum kinetic equation for the density matrix $\rho_{nm}(\bm{k},t)$ will maintain this covariance such that the density $\text{Tr} \hat{\rho}(t)$ and the current density $\text{Tr}\hat{\rho}(t)e\hat{\bm{v}}$ are invariant under these gauge transformations.

\section{Current Densities First Order in an External Electric Field}

Charge currents that are linearly proportional to the electric field can be found by solving equation \ref{mastereq3} for the density matrix to first order in the electric field.  First we expand the density matrix in powers of the electric field $\rho_{nm}(\bm{k})=\sum_p\rho^{(p)}_{nm}(\bm{k})$ where $p$ indexes the order to which $\rho^{(p)}_{nm}(\bm{k})$ is proportional to $\bm{E}(t)$.  For $p=0$ the density matrix is unperturb by the electric field and the solution to equation \ref{mastereq3}  at zeroth order in the external field is just the equilibrium fermi distribution: $\rho_{nm}^{(0)}=\delta_{nm}f^T_n(\bm{k},\mu)$.  The first order equation can now be written as

\begin{equation}\label{firstorder}
\dfrac{\partial \rho^{(1)}_{nm}(\bm{k},t)}{\partial t}+\alpha_{nm}(\bm{k})\rho^{(1)}_{nm}(\bm{k},t)=\dfrac{e\bm{E}(t)\cdot \bm{g}_{nm}(\bm{k})}{\hbar}
\end{equation}

\noindent
where the density matrix to first order in the electric field $\rho^{(1)}_{nm}$ couples to $\alpha_{nm}(\bm{k})=i/\hbar(\varepsilon_n(\bm{k})-\varepsilon_m(\bm{k}))+1/\tau$ and the external perturbing field couples to $g^i_{nm}(\bm{k})=\delta_{nm}\partial_if^T_n(\bm{k},\mu)+i(f^T_n(\bm{k},\mu)-f^T_m(\bm{k},\mu))R_{nm}^i(\bm{k})$.  At zero temperature this coupling leads to two types of terms in the equation of motion for the density matrix.  For $T=0$ terms proportional to $\partial_if^T_n(\bm{k},\mu)$ are only nonzero on the fermi surface as $\partial_if^{T=0}_n(\bm{k},\mu)=\partial_{k_i}\varepsilon_n(\bm{k})\delta(\varepsilon_n(\bm{k})-\mu)$ leading to intraband processes that contribute to $\rho_{nm}(\bm{k},t)$.  The other terms in $g_{nm}^i(\bm{k})$ describe interband processes mediated by the matrix elements of the dipole operator $\bm{R}_{nm}(\bm{k})$.

The solution to equation \ref{firstorder} is

\begin{equation}\label{den1}
\rho_{nm}^{(1)}(\bm{k},t)=\int_{t_p}^tdt'e^{-\alpha_{nm}(\bm{k})(t-t')}\dfrac{e\bm{E}(t')\cdot\bm{g}_{nm}(\bm{k})}{\hbar}
\end{equation}

\noindent
Here we have assumed that the perturbing electric field is zero for times $t<t_p$ ($\bm{E}(t)\sim\theta(t-t_p)$).  Equation \ref{firstorder} demonstrates that indeed $\rho_{nm}^{(1)}(\bm{k},t)$ are matrix elements of a hermitian operator such that taking its trace with respect to $n$, $m$, and $\bm{k}$ or the trace of a product of it and other hermitian operators will lead to quantities whose values are purely real.  The associated current density for example whose values are purely real is found by tracing over the operator product $e\hat{v}\hat{\rho}(t)$:

\begin{equation}
\bm{j}(t)=\dfrac{1}{V}\sum_{k,n,m}e\bm{v}_{nm}(\bm{k})\rho_{mn}(\bm{k},t)
\end{equation}

Here $v^i_{nm}(\bm{k})$ are the matrix elements of the velocity operator in the $\hat{\bm{i}}$-direction written in the Bloch basis.  In general $\hat{\bm{v}}=i[\hat{H},\hat{\bm{r}}]$.  Here we work in {\it length} gauge were the coupling of fermionic matter to the external electric field can be written as $e\bm{E}(t)\cdot \bm{r}$.  For this electromagnetic gauge choice the velocity operator is $\hat{\bm{v}}=i[\hat{H}_0,\hat{\bm{r}}]$ and independent of the electric field.  Here we choose the gauge on the Bloch states $\ket{u_n(\bm{k})}$ such that the velocity operator takes the representation 

\begin{equation}
v^i_{nm}(\bm{k})=\dfrac{1}{\hbar} \bra{u_n(\bm{k})}\partial_{k_i}\hat{H}_0(\bm{k})\ket{u_m(\bm{k})}
\label{vop}
\end{equation}

\noindent
If we had decided to work in {\it velocity} gauge where the coupling to the external electromagnetic potential is described by the perturbation $\hat{H}'(\bm{k})\sim \bm{v}(\bm{k})\cdot \bm{A}(t)$ the velocity operator would contain terms proportional to the external perturbing field \cite{PhysRevB.96.035431,PhysRevB.97.235446}.  In this work we choose to work in {\it length} gauge were the velocity operator is independent of the perturbing electric field and is simply given by equation \ref{vop}. 

Usually one is interested in systems perturbed by electric fields that are oscillatory in time with a single frequency $\omega$.  In these cases it is usually advantageous to look at the Fourier transform $\bm{\mathcal{J}}(\omega)=\int dt e^{-i\omega t}\bm{j}(t)$ of the current density $\bm{j}(t)$.  Equation \ref{den1} is in the form of a convolution such that the Fourier transform of $\bm{j}(t)$ to first order in the electric field is simply

\begin{equation}
\bm{\mathcal{J}}^{(1)}(\omega)=\dfrac{1}{V}\sum_{k,n,m}\dfrac{e^2\bm{E}(\omega)\cdot\bm{g}_{nm}(\bm{k})}{\hbar\alpha_{nm}-i\hbar\omega}\bm{v}_{mn}(\bm{k})
\end{equation}

\noindent
where again $\alpha_{nm}(\bm{k})=i/\hbar(\varepsilon_n(\bm{k})-\varepsilon_m(\bm{k}))+1/\tau$.  

Here we are interested in electric field pulses that are short compared to all other time scales of our system.  We thus consider an electric field pulse $\bm{E}(t)=\bm{E}_0 \Delta_t \delta(t-t_0)$.  For this type of perturbing field the first order contribution to the current is

\begin{equation}\label{cur1}
\bm{j}^{(1)}(t)=\dfrac{\theta(t-t_0)}{V}\sum_{k,n,m}e^{-\alpha_{nm}(t-t_0)}\dfrac{e^2\bm{E}_0\cdot\bm{g}_{nm}(\bm{k})}{\hbar}\Delta_t\bm{v}_{mn}(\bm{k})
\end{equation}

\noindent
The current decays exponentially in time ($\bm{j}^{(1)}(t)\sim e^{-(t-t_0)/\tau}$).  At zero temperature intraband contributions proportional to the diagonal part of the velocity matrix $\bm{v}_{nm}(\bm{k})$ on the Fermi surface are non-osscilatory as $\alpha_{nn}(\bm{k})$ is purely real, while interband contributions between bands $n$ and $m$ oscillate with frequency $(\varepsilon_n(\bm{k})-\varepsilon_m(\bm{k}))/\hbar$.  These interband contributions are summed incoherently across all crystal momentum leading to smooth behavior of $\bm{j}^{(1)}(t)$ for all time $t>t_p$.

\section{Current Densities Second Order in an External Electric Field}

The current density to second order in a perturbing electric field can be found by solving equation \ref{mastereq3} for the density matrix to second order in the perturbation.  Similar to the previous section we first expand the density matrix in powers of the electric field and equate terms on the left and right hand side of equation \ref{mastereq3}   that are quadratically proportional to the perturbation.  This leads to a second order equation for $\rho^{(2)}_{nm}(\bm{k},t)$

\begin{widetext}
\small
\begin{equation}\label{mastereq2}
\dfrac{\partial \rho^{(2)}_{nm}(\bm{k},t)}{\partial t}+\bigg(\dfrac{i}{\hbar}(\varepsilon_n(\bm{k})-\varepsilon_m(\bm{k}))+\dfrac{1}{\tau}\bigg)\rho^{(2)}_{nm}(\bm{k},t)=\sum_{i,l}\dfrac{eE_i(t)}{\hbar}\bigg(\partial_{k_i}\rho^{(1)}_{nm}(\bm{k},t)-i(R^i_{nl}(\bm{k})\rho^{(1)}_{lm}(\bm{k},t)-\rho^{(1)}_{nl}(\bm{k},t)R^i_{lm}(\bm{k}))\bigg)
\end{equation}
\normalsize
\end{widetext}

With knowledge of $\rho^{(1)}_{nm}(\bm{k},t)$ we can use this equation to solve for $\rho^{(2)}_{nm}(\bm{k},t)$.  The solution can be broken into three parts

\begin{widetext}
\begin{align}\label{den2}
\rho^{(2)}_{nm}(\bm{k},t)=\int_{t_p}^tdt''\int_{t_p}^{t''}dt'\dfrac{e^2}{\hbar^2}\sum_{ij}E_j(t'')E_i(t')
e^{-\alpha_{nm}(\bm{k})(t-t'')} (\chi^1_{ij,nm}(\bm{k},t',t'')+\chi^2_{ij,nm}(\bm{k},t',t'')+\chi^3_{ij,nm}(\bm{k},t',t''))
\end{align}
\end{widetext}

\noindent
Here the tensors $\chi^p_{ij}(\bm{k},t',t'')$ each contribute uniquely to the quantum density matrix.

\begin{align}
\chi_{ij,nm}^1(\bm{k},t',t'')&=\partial_{k_j}\alpha_{nm}(\bm{k})(t'-t'')e^{-\alpha_{nm}(t''-t')}g_{nm}^i(\bm{k}) \nonumber \\
\chi^2_{ij,nm}(\bm{k},t',t'')&=e^{-\alpha_{nm}(\bm{k})(t''-t')}\partial_{k_j}g^i_{nm}(\bm{k}) \nonumber\\
\chi^3_{ij,nm}(\bm{k},t',t'')&=\sum_l\bigg(-iR_{nl}^j(\bm{k})g^i_{lm}(\bm{k})e^{-\alpha_{lm}(\bm{k})(t''-t')} \nonumber \\
&+ig^i_{nl}(\bm{k})R_{lm}^j(\bm{k})e^{-\alpha_{nl}(\bm{k})(t''-t')}\bigg)
\end{align}

To demonstrate the solution to these equations for short electric field pulses we again use $\bm{E}(t)=\bm{E}_0\Delta_t\delta(t-t_0)$.  Integration in equation \ref{den2} over this field leads to $t'\rightarrow t_0$ and $t''\rightarrow t_0$.  The first contribution to the second order density vanishes as $\chi^1_{ij,nm}(\bm{k},t_0,t_0)=0$.  This leads to the current density

\begin{align}
\bm{j}^{(2)}(t)=\dfrac{\theta(t-t_0)}{V}\sum_{n,m,i,j}\dfrac{e^3}{\hbar^2}e^{-\alpha_{nm}(\bm{k})(t-t_0)}\bm{v}_{mn}(\bm{k})E_0^iE_0^j\Delta_t^2 \nonumber \\
\times \bigg(\partial_jg_{nm}^i(\bm{k})-i(R_{nl}^j(\bm{k})g_{lm}^i(\bm{k})-g_{nl}^i(\bm{k})R_{lm}^j(\bm{k}))\bigg)
\end{align}

\noindent
At zero temperature we can further divide this response into two pieces $\bm{j}(t)=1/V\sum_k(\bm{j}^{intra}(\bm{k},t)+\bm{j}^{inter}(\bm{k},t))$.  Here $\bm{j}^{intra}(\bm{k},t)$ is nonzero only at crystal momentum on the Fermi surface, while $\bm{j}^{inter}(\bm{k},t)$ has support across the Brillouin zone.  This division can be done uniquely once demanding that each contributions be itself gauge invariant \cite{foldi2017gauge}.  With this constraint

\begin{widetext}

\hspace{10cm}

\begin{gather}
\bm{j}^{inter}(\bm{k},t)=\sum_{n,m,i,j}\dfrac{e^3}{\hbar^2}E_0^iE_0^j\Delta_t^2 e^{-\alpha_{nm}(\bm{k})(t-t_0)}\bigg((f_n^{T=0}(\bm{k},\mu)-f_m^{T=0}(\bm{k},\mu))i\partial_{k_i}R_{nm}^j(\bm{k})v_{mn}^p(\bm{k}) \nonumber\\
+\sum_l(2f_l^{T=0}(\bm{k},\mu)-(f_n^{T=0}(\bm{k},\mu)+f_m^{T=0}(\bm{k},\mu)))R_{nl}^i(\bm{k})R_{lm}^j(\bm{k})v_{mn}^p(\bm{k})\bigg)
\end{gather}

\begin{gather}
\bm{j}^{intra}(\bm{k},t)=\sum_{n,m,i,j}\dfrac{e^3}{\hbar}E_0^iE_0^j\Delta_t^2
\delta(\varepsilon_n-\mu)v_{nn}^i(\bm{k}) \nonumber \\
\times\bigg(-\delta_{nm}\partial_j v_{nm}^p(\bm{k})e^{-\alpha_{nm}(\bm{k})(t-t_0)}
+2i(e^{\alpha_{nm}(\bm{k})(t-t_0)}R^j_{nm}(\bm{k})v^p_{mn}(\bm{k})-e^{-\alpha_{mn}(\bm{k})(t-t_0)}v^p_{nm}(\bm{k})R^j_{mn}(\bm{k}))\bigg)
\end{gather}
\end{widetext}

\noindent
Both contributions decay exponentially in time with timescale $\tau$.  The interband contributions also oscillate with frequencies determined by the energy differences between bands.  This incoherent summation over the entire Brillouin zone leads to a contribution to the current smooth in time.  The intraband contributions have two parts.  One part is proportional to diagonal elements of $\alpha_{nm}(\bm{k})$ leading to non-oscillatory contributions to the current.  The other part oscillates with frequency again defined by the energy difference between bands.  We will show in the next section that for certain band structures the sum of these terms across the Fermi surface can add coherently when the energy differences between band states on the Fermi surface and the states just above the Fermi surface are nearly constant.  This coherent superposition of terms that oscillate at a fixed frequency can lead to currents that oscillate in time with frequency determined by these energy differences as we will now demonstrate.

\section{Minimal Models for Oscillatory Charge Current Generation}

In the previous section we proved that the induced charge current to second order in an ultrafast electric field pulse has two contributions: an interband contribution that develops currents that are smooth in time and an intraband piece that for the right model can develop currents that oscillate in time.  Here we demonstrate this phenomena in a minimal two band model.

The model consists of a continuum theory of two valleys that can represent the low energy dynamics of spin-less electrons in a two dimensional crystal.  The Bloch hamiltonian in an orbital basis takes the form

\begin{equation}\label{hamil}
\hat{H}_{\chi}(\bm{k})=\chi \hbar \bm{b}\cdot\bm{k} \mathbb{I}+\hbar v_F (\chi k_x \sigma_x +k_y \sigma_y) +m_0 \sigma_z
\end{equation}

\begin{figure*}
  \includegraphics[width=\linewidth]{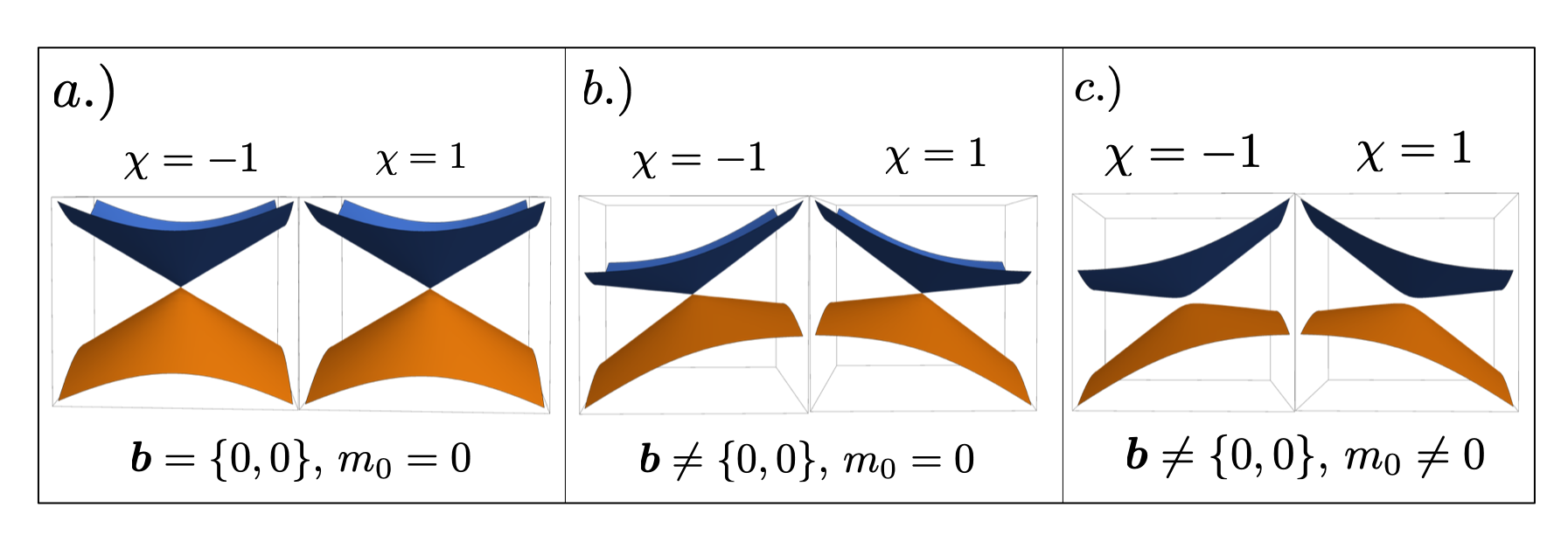}%
\caption{(a)-(c) Band structures of various two band models described by equation \ref{hamil}.  (a) Model preserving time reversal, inversion, and chiral symmetries.  (b) Model with time reversal and inversion symmetries, but with broken chiral symmetry.  (c) Model with time reversal symmetry, but broken inversion and chiral symmetries.} \label{bandstruc}
\end{figure*}

\noindent
Here $\chi=\pm1$ indexes the valley degree of freedom.  For spin-less electrons time reversal $\mathcal{T}$ is just the complex conjugation operator $\mathcal{K}$.  We imagine that the valleys are centered at opposite crystal momentum in the Brilloiun zone such that the Hamiltonian is time reversal invariant and satisfies $H^*_{1}(\bm{k})=H_{-1}(-\bm{k})$. For vanishing $m_0$ and $\bm{b}$ the theory consists of linear bands that cross at $\bm{k}_0=(0,0)$ for each valley (Figure \ref{bandstruc}a).  In this limit the hamiltonian has a chiral symmetry $\{\hat{H}(\bm{k}),\sigma_z\}=0$ such that its energy eigenvalues come in plus/minus pairs.  Also in this limit the hamiltonian satisfies $\sigma_x\hat{H}_1(\bm{k})\sigma_x=\hat{H}_{-1}(-\bm{k})$ and has inversion symmetry.  Nonzero $\bm{b}$ tilts the linear bands such that the chiral symmetry is broken and the spectrum at each valley no longer consists of plus/minus pairs (Figure \ref{bandstruc}b).  Nonzero $m_0$ breaks inversion and introduces a gap that breaks the two fold degenerate crossing at $\bm{k}_0$ for each valley (Figure \ref{bandstruc}c).

\begin{figure}
  \includegraphics[width=.8\linewidth]{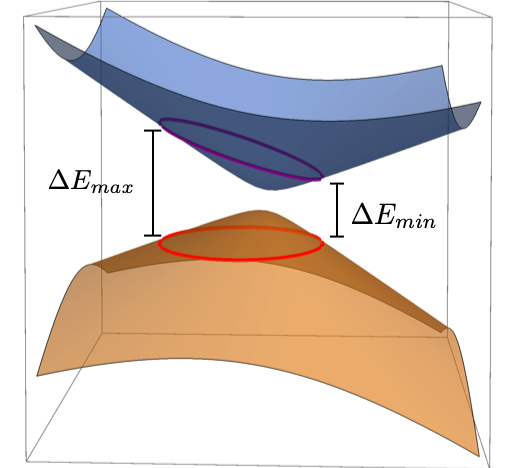}%
\caption{Typical time reversal invariant, inversion and chiral breaking band structure for a single valley $\chi$.  Fermi surface is schematically shown in red.  Contributions to $\bm{j}^{intra}(\bm{k})$ derive from interband matrix elements between states. on the Fermi surface and Bloch states along the purple ellipse.  Terms in $\bm{j}^{intra}(\bm{k})$ oscillate at frequencies $\Delta E_{min}/\hbar\leq\omega\leq\Delta E_{max}/\hbar$.} \label{transitions}
\end{figure}

For our ultrafast electric field pulses $\bm{E}(t)=\bm{E}_0 \Delta_t \delta(t-t_0)$ the intraband contribution to the current will be nonzero if both chiral and inversion symmetries are broken and at electron filings for which the chemical potential sits below the energy gap.  In these situations the Fermi surface will be an ellipse in the $k_xk_y$-plane.  Contributions to $\bm{j}^{intra}(\bm{k})$ come from transitions between these Fermi surface states and the states in the unoccupied band with the same Bloch momenta.  These terms each have a contribution that oscillate at a frequency determined by the energy difference between these bands at momenta along the Fermi surface.  For vanishing $\bm{b}$ the energy difference would be constant across the Fermi surface and equal to $2\mu$.  This would lead to terms in $\bm{j}^{intra}(\bm{k})$ that oscillate at frequency $\omega=2\mu/\hbar$.  Nonzero $\bm{b}$ leads to a distribution of energy differences across the Fermi surface.  Figure \ref{transitions} shows the band structure in a single valley for a typical time reversal invariant, but chiral and inversion broken system.  The Fermi surface is schematically shown in red.  States with momenta along the Fermi surface, but in the unoccupied band are shown in purple.  The terms $\bm{j}^{intra}(\bm{k})$ contributing to the intraband contribution to the current will oscillate over a range of frequencies: $\Delta E_{min}/\hbar\leq\omega\leq\Delta E_{max}/\hbar$, where $\Delta E_{max}$ and $\Delta E_{min}$ depend on $m_0$, $\bm{b}$, and $\mu$.

In order to observe an oscillatory current generated from an ultrafast electric field pulse the terms in $\bm{j}^{intra}(\bm{k})$ must add coherently and thus we must be in the limit where $(\Delta E_{max}-\Delta E_{min})/\hbar<<\Delta\bar{E}$, where $\Delta \bar{E}$ is the average energy difference between the two bands across the Fermi surface.  In this regime these interband matrix coefficients will lead to a current that oscillates with frequency $\bar{\omega}=\Delta \bar{E}/\hbar$.  To tune this frequency one can adjust the chemical potential of the system thereby changing the average energy difference between the two bands around crystal momentum along the Fermi surface and thus changing $\bar{\omega}$.  Furthermore the current generated from these ultrafast electric field pulses decays exponentially in time with timescale $\tau$.  To measure multiple periods of oscillation $2\pi/\bar{\omega}<<\tau$.

\begin{figure}
\vspace{.5cm}
\sidesubfloat[]{
  \includegraphics[width=.8\linewidth]{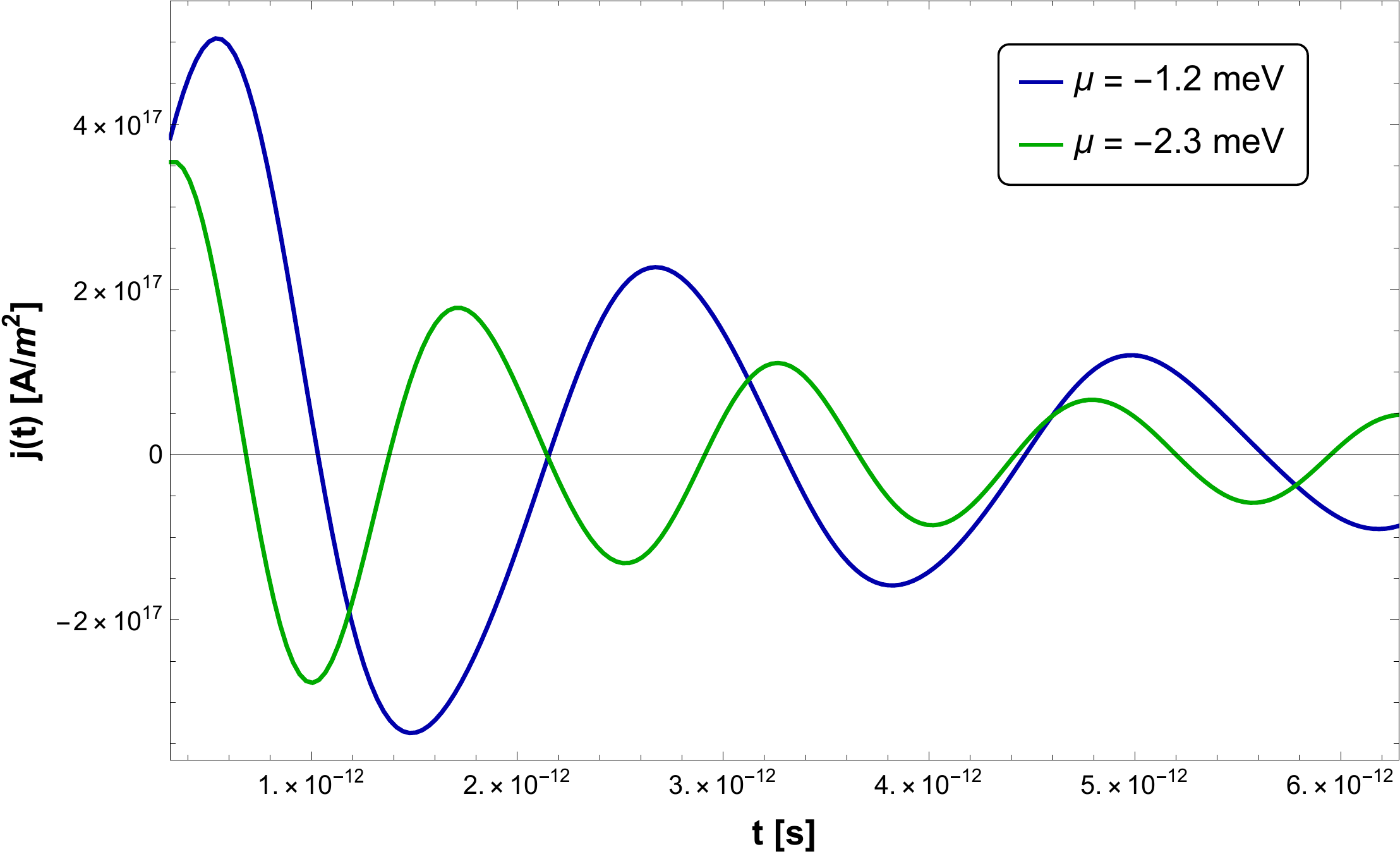}} \\ \vspace{.4 cm}%
  \sidesubfloat[]{
    \includegraphics[width=.8\linewidth]{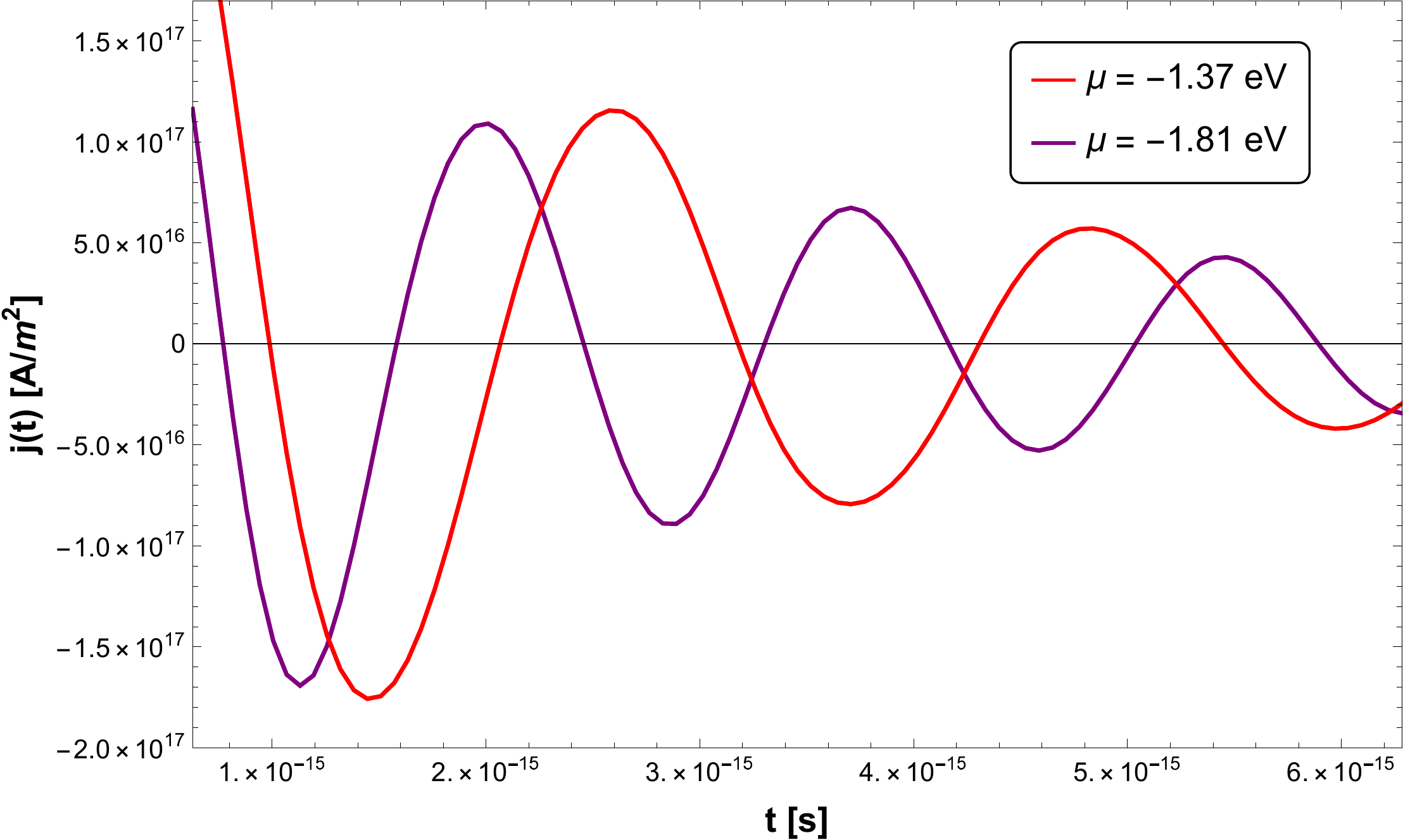}}%
\caption{The intraband contribution to the current as a function of time for a system with $\bm{b}=(0.5,0.2) v_f$, $v_f=10^6$ m/s, and $\tau=6.3\times10^{-15}$ s perturbed by an electric field pulse with $\Delta_t=0.2$ ps and $E_0=5$ MV/cm. (a) System with small gapsize $m_0=0.001 \hbar v_f/a$ for two different chemical potentials $\mu=-1.21$ meV and $\mu=-2.3$ meV with average terahertz frequency modulation $\bar{\omega}=2.79\times 10^{12}$ s$^{-1}$ and $\bar{\omega}=4.21\times10^{12}$ s$^{-1}$ respectively. (b) System with large gapsize $m_0=0.5 \hbar v_f/a$ for two different chemical potentials $\mu=-1.37$ eV and $\mu=-1.81$ eV with average petahertz frequency modulation $\bar{\omega}=2.86\times 10^{15}$ s$^{-1}$ and $\bar{\omega}=3.69\times10^{15}$ s$^{-1}$ respectively. } \label{timej}
\end{figure}

Figure \ref{timej} shows the intraband contribution to the current for a system with $\bm{b}=(0.5,0.2) v_f$, $v_f=10^6$ m/s, and $\tau=6.3\times10^{-15}$ s perturbed by an electric field pulse with $\Delta_t=0.2$ ps and $E_0=5$ MV/cm.  In Figure \ref{timej}(a) the system has a small gapsize with $m_0=0.001 \hbar v_f/a$.  The intraband contribution to the current as a function of time is plotted for this system at two different chemical potentials $\mu=-1.21$ meV and $\mu=-2.3$ meV.   The currents are shown to modulate in time with mean terahertz frequencies $\bar{\omega}=2.79\times 10^{12}$ s$^{-1}$ and $\bar{\omega}=4.21\times10^{12}$ s$^{-1}$ respectively.  In figure \ref{timej}(b) the system has a large gapsize $m_0=0.5 \hbar v_f/a$.  Again the intraband contribution to the current is shown for two different chemical potentials $\mu=-1.37$ eV and $\mu=1.81$ eV.   The average petahertz frequency modulation of the currents are $\bar{\omega}=2.86\times 10^{15}$ s$^{-1}$ and $\bar{\omega}=3.69\times10^{15}$ s$^{-1}$ respectively.  These examples demonstrate the robustness to generate coherent oscillating currents at frequencies from terahertz all the way to petahertz by manipulation of a chemical potential.

\section{Conclusion}

Here we have demonstrated that ultrafast electronic field pulses can induce currents nonlinear in the perturbing electric field that modulate at a frequency determined by the energy differences between bands.  These oscillating currents arise from interband transitions from electronic states on the Fermi surface to unoccupied states with equal crystal momentum.  Other contributions to the current derive from matrix elements located across the Brillouin zone that oscillate with a wide range of frequencies and ultimately lead to an incoherent superposition of terms that result in non-oscillatory current behavior.  The frequency of the intraband contribution to the current can be manipulated by changing the chemical potential which tunes the average energy difference between states on the Fermi surface and states above it.  This tool in principle can be  used as a mechanism for generating currents that oscillate at various frequencies from THz to the near IR.

Once an induced current is generated it can radiate into the outgoing solutions of the Maxwell wave equation

\begin{equation}
j^{ind}_i(\bm{r},t)=\sum_j\dfrac{1}{\mu_0}\bigg(\delta_{ij}\dfrac{1}{c^2}\dfrac{\partial^2}{\partial t^2}-\bm{\nabla}\cdot\bm{\nabla}+\partial_{r_i}\partial_{r_j}\bigg)A^{ind}_j(\bm{r},t)
\end{equation}

\noindent
Induced charge currents at frequency $\omega$ will generate induced electromagnetic fields $\bm{A}^{ind}(\bm{r},t)$ at the same frequency.

For high frequency current modulations, measurement of these induced electromagnetic fields can be done through techniques like attosecond interferometry \cite{mashiko2018multi}.  Studying the high frequency electronic dynamics in materials creates a new platform for ultrafast electronic logic and signal processing.  Identifying a precise mechanism for their generation now allows for the development of new control and manipulation protocols implemented in the contexts of semiconductor band engineering.

\section{Acknowledgments}

ZA and EJM are supported by the Department of Energy under Grant No. DE-FG02-84ER45118.


\nocite{*}
\bibliography{main}

\end{document}